\documentclass[twoside]{ilcws08}
\usepackage[latin1]{inputenc}
\usepackage[dvips]{}
\usepackage{wrapfig,rotating}
\usepackage{amssymb,amsmath,array}

\usepackage{tabularx,colortbl}

\begin{document}
\title{A digital ECAL based on MAPS}

\author{J.~A.~Ballin, P.~D.~Dauncey, A.-M.~Magnan$^\dagger$, M.~Noy$^1$\\
	Y.~Mikami, O.~Miller, V.~Rajovi\'{c}, N.~K.~Watson, J.~A.~Wilson$^2$\\
	J.~P.~Crooks, M.~Stanitzki, K.~D.~Stefanov, R.~Turchetta,
      M.~Tyndel, E.~G.~Villani$^3$ 
\vspace{.3cm}\\
1 - Department of Physics, Imperial College London, London, UK\\
2 - School of Physics and Astronomy, University of Birmingham, Birmingham, UK\\
3 - Rutherford Appleton Laboratory, Chilton, Didcot, UK\\
$\dagger$ - Contact: {\tt A.Magnan@imperial.ac.uk}
}

\maketitle

\begin{abstract}
Progress is reported on the development and testing of Monolithic
Active Pixel Sensors (MAPS) for a Si-W ECAL for the ILC. Using laser
and source setups, a first version of the sensor has been
characterised through measurements of the absolute gain calibration,
noise and pedestal. The pixel-to-pixel gain spread is 10\%. 
Charge diffusion has been measured and found to be compatible with
simulation results. The charge collected by a single pixel varies from
50\% to 20\% depending on where it is generated. After adding detector
effects to the Geant4 simulation of an ILC-like ECAL, using the
measured parameters, the energy resolution is found to be 35\% higher
than the ideal resolution, but is still lower than the resolution
obtained for an equivalent analogue ECAL.

\end{abstract}

\section{Motivation}
\label{sec:motiv}

\begin{wrapfigure}{r}{0.4\columnwidth}
\vspace*{-1.cm}
\centerline{\includegraphics[width=0.4\columnwidth,height=0.4\columnwidth]{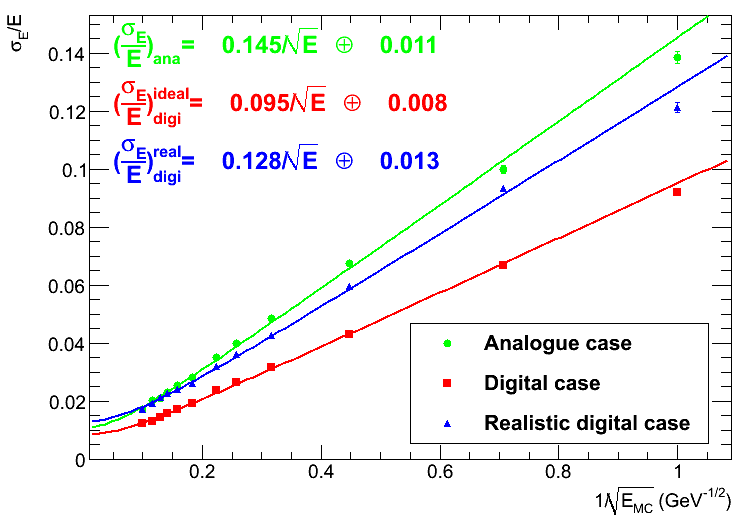}}
\caption{Energy resolution as a function of the incident energy for
single electrons, for both analogue and digital approaches (see
Section~\ref{sec:phys} for more details). The three formulae given
correspond to the fitted functions displayed.}
\label{Fig:resIdeal}
\end{wrapfigure}
In a sampling electromagnetic calorimeter (ECAL), the energy deposited
is proportional to the number of charged particles created in the
shower, itself proportional to the incident energy of the
particle. Two approaches are hence possible to measure the incident
energy: by measuring the number of charged particles (digital
approach) or by measuring the energy deposited (analogue
approach). The motivation for using digital over analogue lies in the
fluctuations occurring in the measured quantity. Only fluctuations in
the development of the shower are expected if we are able to truely
count charged particles. The energy deposited is however subject to
additional fluctuations~\cite{slides}, among which the dominant one is
due to the Landau spread. The ideal energy resolution for both
approaches, obtained with a Geant4~\cite{geant4} simulation of an
ILC-like ECAL (20 layers at 0.6\,X$_0$ followed by 10 layers at
1.2\,X$_0$, and 500\,$\mu$m silicon thickness per layer)~\cite{ILD},
is shown in Figure~\ref{Fig:resIdeal}. The digital approach is indeed
about 30\% lower in energy resolution than the analogue approach. From
ideal to real conditions, the analogue resolution is however expected
to change very little, whereas in the digital case, it is subject to
charge diffusion and more importantly depends on the degree to which
the number of charged particles can be measured. The following
question needs to be answered: how close can we approach the ideal
resolution for the digital case?


\section{Sensor testing}
\label{sec:dev}

A prototype sensor of $1 \times 1$\,cm$^2$ with a pixel size of $50
\times 50$\,$\mu$m$^2$ was fabricated in 2007, with four different
pixel alternatives. The one found to have the best performance
(preShape architecture~\cite{TPAC}) is described in this section. Each
pixel is made of four diodes, situated near the four corners,
connected to a charge preamplifier and CR-RC shaper, and a two-stage
comparator with individual threshold settings. When a particle
generates a signal above the configured threshold, the spatial
coordinates and time-stamp are recorded in a 13-bit word. Due to the
timing expected at the ILC~\cite{ILD}, the data must be stored on the
sensor for the entire bunch train ($\simeq 1$\,ms or 2600 bunch
crossings), and then read out during the 200\,ms available between
bunch trains. To reduce the impact of the dead area needed for memory
and logic, these have been placed in columns 5 pixels wide
(250\,$\mu$m) every 42 active pixels.

To increase the charge collection efficiency, a deep p-well implant
has been added (INMAPS process) to the standard 0.18\,$\mu$m process
used, to shield the epitaxial layer from the electronics n-wells. Its
performance is reported in Section~\ref{sec:chargeDiff} and more
details can be found in~\cite{TPAC}.

The characterisation of the sensor has been made through three main
measurements. For each, the general method involves threshold scans,
in ``Threshold Units'' (TU), as the readout is binary. Each time a
threshold scan is recorded with an input signal (source/laser), a
corresponding noise-only threshold scan needs to be taken
systematically for comparison.


\subsection{Absolute signal calibration}

\begin{wrapfigure}{r}{0.4\columnwidth}
\vspace*{-1.5cm}
\centerline{\includegraphics[width=0.38\columnwidth]{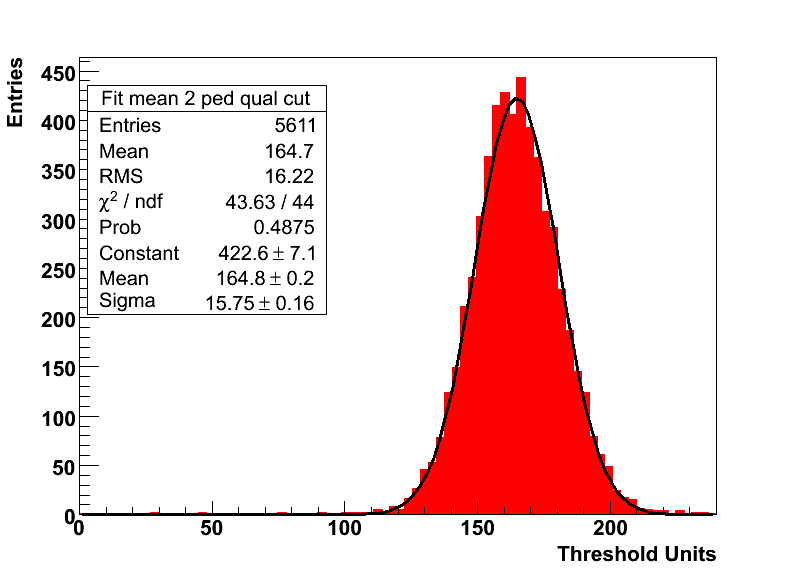}}
\caption{Gain measurement from a $^{55}$Fe source for all pixels studied.}\label{Fig:Fe55gain}
\end{wrapfigure}
The absolute signal calibration is studied using a $^{55}$Fe
source. The emitted photons deposit all their energy, 5.9\,keV or
$\simeq$ 1620\,electrons, within 1\,$\mu$m$^3$ of silicon. By
differentiating the threshold scan spectrum obtained per pixel, a peak
can be observed whose central value should correspond to the total
charge of 1620\,electrons or 5.9\,keV. The results for all studied
pixels are shown in Figure~\ref{Fig:Fe55gain}. The gains of the pixels
are found to be uniform within 10\%. The conversion factors obtained
are 1\,TU $\simeq$ 10\,e$^{-} \simeq$ 36\,eV.

\subsection{Pedestal and noise measurements}

\begin{wrapfigure}{r}{0.4\columnwidth}
\vspace*{-1.5cm}
\centerline{\includegraphics[width=0.38\columnwidth]{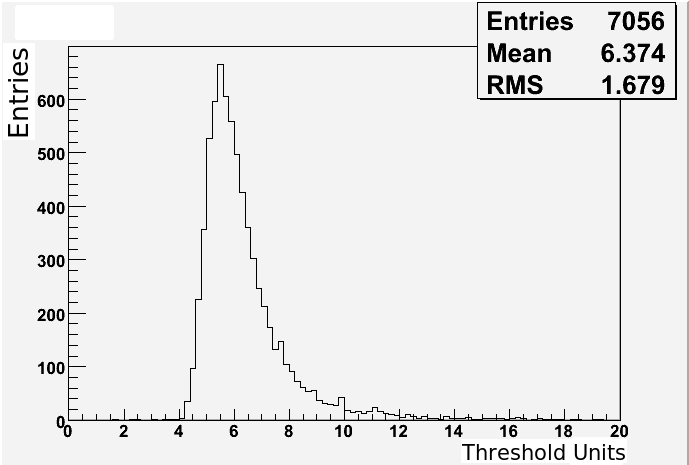}}
\caption{Noise distribution for all studied pixels.}\label{Fig:noise}
\end{wrapfigure}
Noise and pedestal are defined by the mean and RMS in single-pixel
noise-only threshold scans. By enabling only one pixel at a time,
crosstalk effects are avoided. On average, the noise is 6\,TU
(60\,e$^-$ or 220\,eV), with a minimum value of 4\,TU (40\,e$^-$ or
140\,eV) as shown in Figure~\ref{Fig:noise}, and no correlation is
found with the position on the sensor. If no trimming is applied on
the individual pixels, the distribution of pedestals for all pixels
studied
shows a spread of about four times the single-pixel noise. By trimming
each pixel using the 4-bit adjustable settings, the spread can be
reduced to the size of the single-pixel noise.




\subsection{Charge sharing measurement}
\label{sec:chargeDiff}

A laser was scanned across the pixel to measure the variation in
charge collected.
The laser spot-size is about 2\,$\mu$m, and its wavelength is
1064\,nm. Silicon is transparent at this wavelength so the laser is
fired at the back plane and focused on the epitaxial layer. 

\begin{wrapfigure}{r}{0.5\columnwidth}
\begin{minipage}[h]{0.24\columnwidth}
\centerline{\includegraphics[width=0.95\columnwidth]{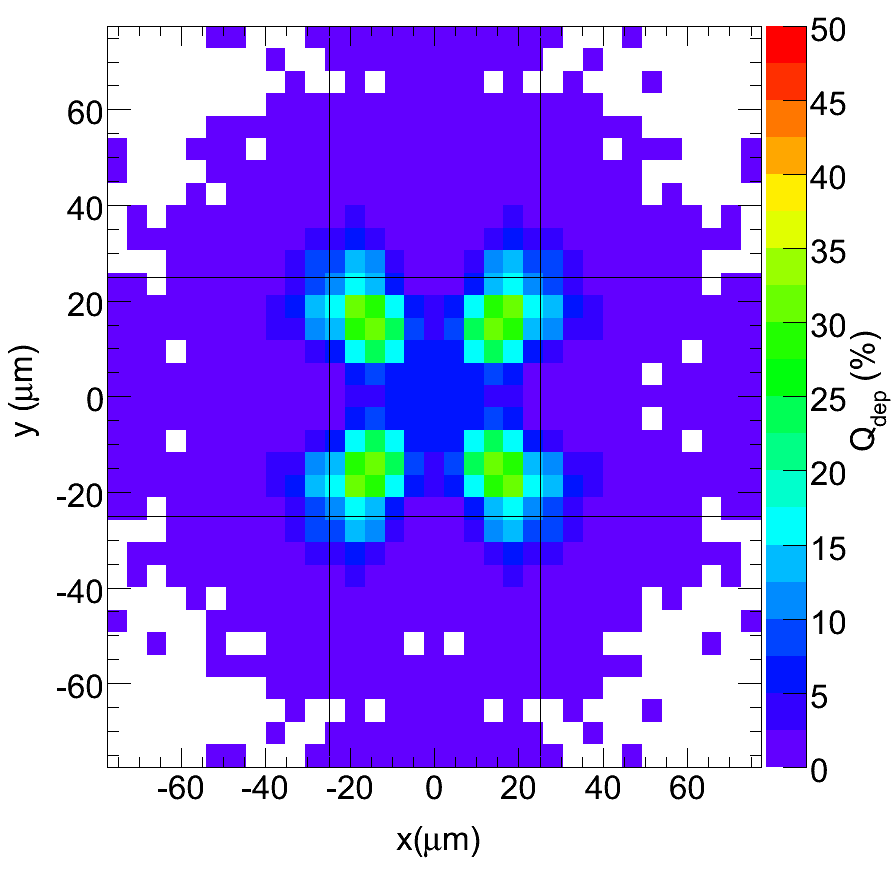}}
(a)
\end{minipage}
\hfill
\begin{minipage}[h]{0.24\columnwidth}
\centerline{\includegraphics[width=0.95\columnwidth]{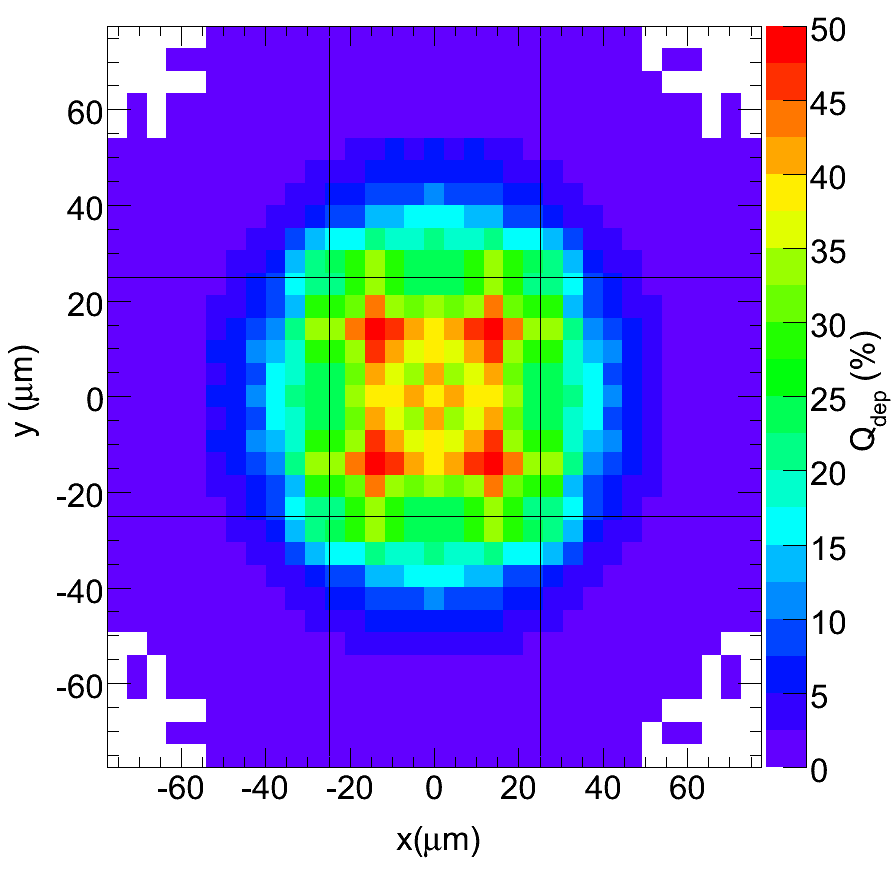}}
(b)
\end{minipage}
\hfill
\begin{minipage}[h]{0.24\columnwidth}
\centerline{\includegraphics[width=0.95\columnwidth]{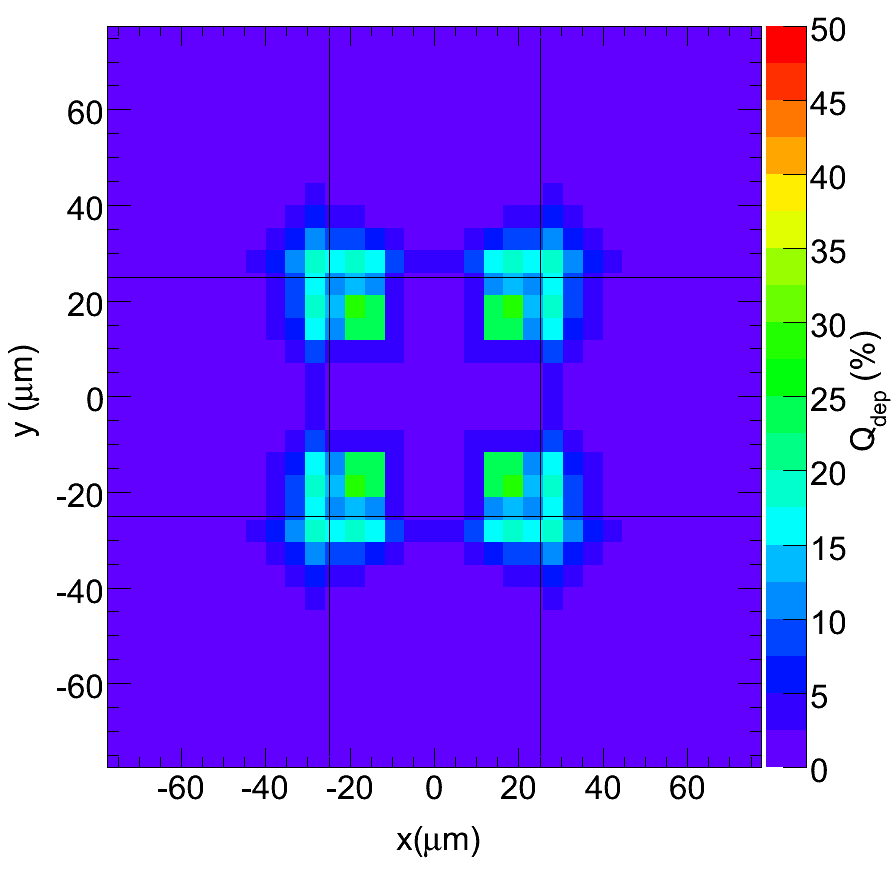}}
(c)
\end{minipage}
\hfill
\begin{minipage}[h]{0.24\columnwidth}
\centerline{\includegraphics[width=0.95\columnwidth]{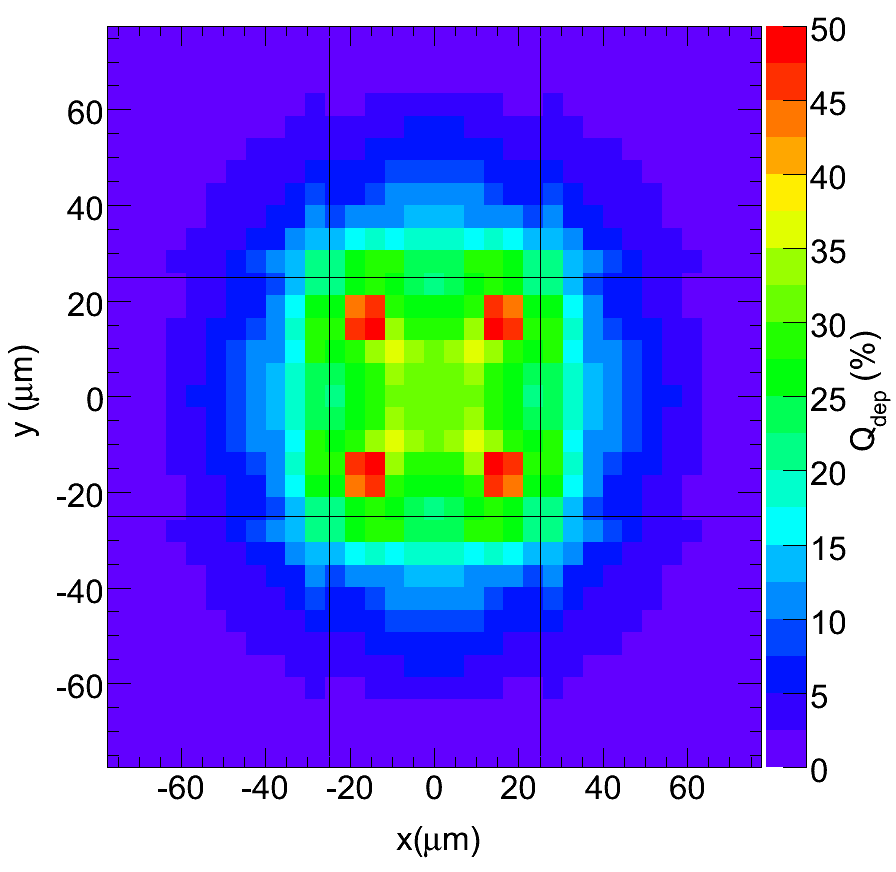}}
(d)
\end{minipage}
\caption{Percentage of charge recorded in a pixel as a function of the
laser input position, for real data (top) and simulation (bottom),
without (left) and with (right) deep p-well implant.
}\label{Fig:chargeDiff}
\end{wrapfigure}
A pixel is scanned in 5\,$\mu$m steps, and threshold
scans are recorded for each step. The charge deposited is found
by fitting the falling edge of each spectrum. In order to demonstrate
the effect of the deep p-well implant, two sensors are compared, one
produced with the INMAPS process and the other without. Results for
real data are shown in Figure~\ref{Fig:chargeDiff}(a) and
\ref{Fig:chargeDiff}(b).

A simulation is performed in similar conditions using the Sentaurus
TCAD software~\cite{TCAD}. Results are shown in
Figure~\ref{Fig:chargeDiff}(c) and \ref{Fig:chargeDiff}(d).  The
simulation reproduces the data quite well, with similar level of
maximum and minimum charge collected for each sensor. The effect of
the deep p-well implant is also confirmed, with a factor 2 to 4
increase in the charge collected. The maximum charge is measured as
expected for deposits near the diodes, at 50\% of the input
charge. The minimum charge, for a deposit inside the pixel, is
measured at the four corners and in the middle part of the four sides,
at 20\% of the input charge. Note, the maximum charge collected from a
corner would be 25\%, from symmetry.

\section{Physics expectations}
\label{sec:phys}

\begin{wraptable}{r}{0.45\columnwidth}
\vspace*{-1.5cm}
\begin{tabular}{|l|c|}
\hline
Effect & Degradation \\
\hline
Noise $\times 2$ & 5\% \\
Dead area 11\% & 6\% \\
+ Sensor edges 5\% & 2\% \\
Charge diffusion & 5\% \\
MIP counting & 20\% \\
\hline
\end{tabular}
\caption{Percentage of degradation in energy resolution
($\sigma_{E}/E$) for 10\,GeV electrons when varying parameters in the
digitisation procedure.}\label{tab:digi}
\end{wraptable}
Now that the charge spread simulation has been validated, and the
noise measured, more realism can be added to the Geant4~\cite{geant4}
Monte Carlo simulation of a real-size detector. Starting from an
energy deposit in $5 \times 5$\,$\mu$m$^2$ cells, charge sharing is
applied according to the simulation results described in
Section~\ref{sec:chargeDiff}. The energy is then summed in $50 \times
50$\,$\mu$m$^2$ cells. Noise is added to signal hits by smearing the
energy by 120\,eV (32\,e$^-$), the target for future versions of the
sensor. Hits above threshold are recorded. Noise-only hits are added
according to the threshold value. The influence of each effect is
studied on the energy resolution as a function of the threshold, for
10\,GeV electrons. Results are summarised in Table~\ref{tab:digi}.

\begin{wrapfigure}{r}{0.5\columnwidth}
\centerline{\includegraphics[width=0.45\columnwidth]{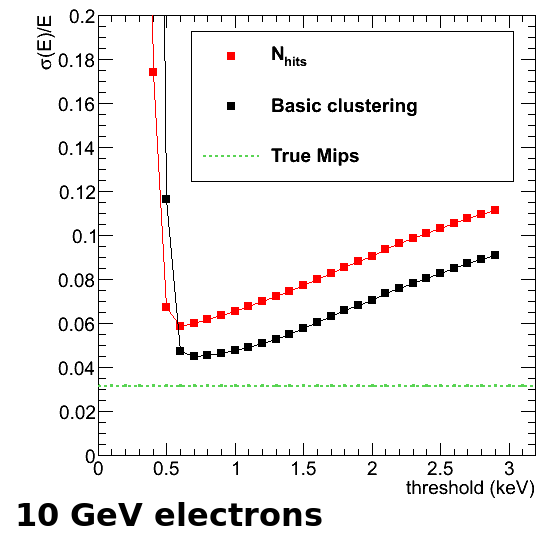}}
\caption{Effect of a MIP-finder algorithm on the energy resolution as
a function of threshold.}\label{Fig:resvsthresh}
\end{wrapfigure}
 The main effect is due to confusion in counting the true number of
MIPs. A simple MIP-finder algorithm has been developed based on
closest neighbours. The result is shown in
Figure~\ref{Fig:resvsthresh}: even after MIP clustering, the
resolution is significantly higher than the ideal case of counting
true MIPs. The MIP-finder algorithm is hence crucial, and needs to be
optimised. However, the Geant4 simulation at such a small pixel size
has never been cross-checked with real data, and beam tests are
required to confirm its validity before any optimisation can be
meaningful.

The energy resolution for a range of energies between 1 and 200\,GeV
is shown in Figure~\ref{Fig:resIdeal} after digitisation. The overall
degradation compared to the ideal case is around 35\%.


\section{Conclusion}
\label{sec:conclu}

A first version of a sensor dedicated to study digital electromagnetic
calorimetry has been developed and characterised using source and
laser setups. The pixel-to-pixel gain spread is 10\%, and the average
noise is 60\,e$^-$ with an RMS of about 20\,e$^-$. Trimming is
required and more trim bits have been added in the second version of
the chip to decrease the pedestal spread below the noise level. Charge
sharing has been measured and simulation is found to reproduce the
real data. The INMAPS process is found to be crucial, doubling
(quadrupling) the maximum (minimum) charge collected. These results
have been applied to a Geant4 simulation of a real-size detector, and
the energy resolution is found to be degraded by about 35\% after
digitisation compared to a true MIP-counting configuration, but still
lower than the corresponding analogue design. The main effect is hit
confusion, but real data are needed to validate the simulation at such
a small pixel size, before MIP-clustering algorithms can be optimised.



\begin{footnotesize}



%

\end{footnotesize}


\end{document}